# To Prolong the Lifetime of Wireless Sensor Networks: The Min-max Fair Scheduling in a Multi-access Power Contra-polymatroid

Xiaomao Mao, Huifang Chen, Peiliang Qiu

Institute of Information and Communication Engineering

Zhejiang University, Hangzhou, 310027 China.

{maoxm, chenhf, qiupl}@zju.edu.cn

## Abstract

From an information-theoretic point of view, we investigate the min-max power scheduling problem in multi-access transmission. We prove that the min-max optimal vector in a contra-polymatroid is the base with the minimal distance to the equal allocation vector. Because we can realize any base of the contra-polymatroid by time sharing among the vertices, the problem searching for the min-max optimal vector is converted to a convex optimization problem solving the time sharing coefficients. Different from the traditional algorithms exploiting bottleneck link or water-filling, the proposed method simplifies the computation of the min-max optimal vector and directly outputs the system parameters to realize the min-max fair scheduling. By adopting the proposed method and applying the acquired min-max optimal scheduling to multi-access transmission, the network lifetime of the wireless sensor network is prolonged.

## I. Introduction

Wireless Sensor Networks (WSNs) are built up by small, inexpensive, and low-power sensor devices with limited processing and communicating capabilities. When properly programmed and deployed in a large scale, such networked sensors can collaborate to accomplish various high level tasks. Since sensor nodes are generally supported by small size batteries

This work was supported by National Natural Science Foundation of China (Grant No.60772093).

whose replacement can be costly if possible, sensor network operations must be energy-efficient in order to maximize the network lifetime.

To prolong the lifetime of WSNs, a lot of researches focus on minimizing the transmission energy under various delay constraints. In [9], [11], [16], energy-efficient transmission scheduling in a point to point (p2p) transmission is considered. These work is motivated by the following observation: It is possible to significantly lower the transmission energy by transmitting the packet over a long period of time. With the arriving time of the packets previously known, Uysal-Biyikoglu et.al proposed the optimal scheduling strategy with minimized energy consumption [16]. In addition, Gamal et.al extended the results to broadcast channel as well as multi-access channel in [2], [15]. However, requiring noncausal packet arriving information and employing iterative algorithms, the proposed strategies are too complicated to implement in a practical network. Other work concerning energy efficiency in WSNs include [1], [5]. The authors considered more practical issues from a cross-layer view and adopted a Time Division Multiple Access (TDMA) transmission scheme. Without fulfilling the bandwidth efficiency, orthogonal access schemes, such as TDMA, are essentially suboptimal in a multi-access channel. Besides, merely decreasing the transmission energy is insufficient to maximize the network lifetime, the fairness of the energy consumption among sensor nodes is a factor of primary importance. To prolong the lifetime of WSNs, we investigate the min-max fair scheduling problem in a general multi-access channel where the users transmit simultaneously upon the entire bandwidth.

With a cluster head or a fusion center as the common receiver and the sensor nodes as the transmitting users, a multiple access channel model is formed in a WSN. The capacity region of a multi-access channel possesses a polymatroid structure, while the power region possesses a contra-polymatroid structure. In [3], [4], [17], the polymatroid and contra-polymatroid structures and their properties were discussed. As there is a duality between polymatroid and contra-polymatroid, they exhibit similar properties. Particularly, the capacity polymatroid and power contra-polymatroid of the multi-access channel were discussed in [12], [14]. A vertex of the polymatroid (contra-polymatroid) can be realized by successive decoding, which means that a series of single-user decodings is sufficient to achieve the vertex. More precisely, first one user is decoded, treating all other users as noise, then its decoded signal is subtracted from the received signal, then the next user is decoded and subtracted, and so forth. Varying the successive decoding order, which is denoted as a permutation on the index set, we can achieve any vertex of the polymatroid (contra-polymatroid). As the dominant face is a convex hull of the vertices, we can realize any base in the dominant face by time sharing among the vertices. In the power contra-polymatroid, the dominant face characterizes the minimum sum power required to implement a given rate vector. Therefore, we will focus on the dominant face and search for the min-max optimal base in the power contra-polymatroid.

As the dominant face is a convex compact set, the min-max optimal vector exists and coincides with the lexicographically optimal vector. In [3], [10], the relation between lexicographically order and min-max fairness was shown. Exploiting this relation and the properties of a contra-polymatroid structure, we first show by theoretical proof that the min-max optimal base is the base with the minimal distance to the equal allocation vector. Furthermore, the min-max optimal base is the min-max optimal vector in the power contra-polymatroid. Because any base of the contra-polymatroid can be realized by time sharing among the vertices, the min-max scheduling problem is transformed to a convex quadratic optimization problem solving the time sharing coefficients. By time sharing according to the obtained coefficients, the min-max fair scheduling is achieved in a multi-access transmission. The main contributions of this paper are summarized as follows:

- From an information-theoretic aspect, we examine the geometry property of the min-max fairness in a contra-polymatroid. We prove that the min-max optimal point in a contra-polymatroid is the point with the minimal distance to the equal allocation point.
- Since any power scheduling strategy can be achieved by time sharing among the successive-decoding-achievable vertices, we convert the min-max scheduling problem to a problem computing the time sharing coefficients. Exploiting the geometry property of the min-max fairness, this problem is formulated as a convex optimization problem and solved by existing algorithms.
- Though analyzed in the framework of multi-access WSN, the proposed method can readily be applied to provide min-max fairness in a wide class of multiuser systems, such as the single-cell uplink scenario where a set of mobiles communicate to the base station with a single receiver.

Dual to the min-max optimal problem considered in this paper, Maddah-Ali et.al discussed the max-min rate allocation in a multi-access capacity polymatroid in [6]–[8]. By decomposing the capacity polymatroid into a set of nested polymatroids, they thoroughly examined the max-min fair scheduling problem in the multi-access capacity polymatroid, including the max-min vertex and the max-min base. However, the proposed algorithms were essentially a tailored version of traditional algorithms in computer networking and utilized similar concepts as bottleneck link or water filling. Besides, the max-min allocation vector and its associated time sharing coefficients can but be acquired separately by recursion. On the contrary, the method proposed in this paper reveals the geometry property of the min-max fairness in a contra-polymatroid. It can compute the min-max fair vector as well as its time sharing coefficients together by adopting an arbitrary convex optimization algorithm. Furthermore, the proposed method is ready to be extended to solve its dual problem for max-min rate allocation in a multi-access capacity polymatroid.

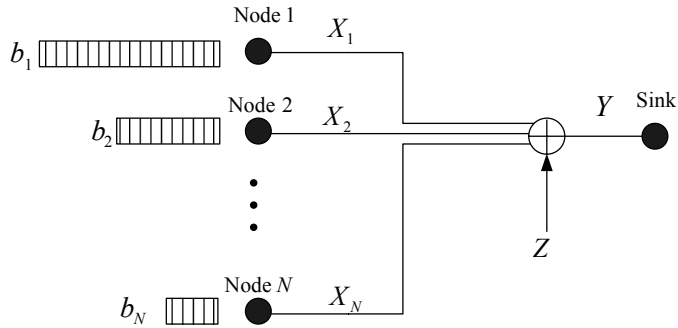

Fig. 1. The model of the multi-access wireless sensor network.

The rest of the paper is organized as follows. We introduce the system model and define the network lifetime in Section II. A review of the polymatroid structure of Gaussian multi-access transmission is followed in Section III. In Section IV, we propose the energy minimized scheduling as a comparison. Section V is the main part of this paper. We prove that the min-max optimal vector in a power contra-polymatroid is the base with the minimal distance to the equal allocation vector. Then, the problem searching for the min-max optimal base is converted to a convex optimization problem computing the associated time sharing coefficients. We present simulation results in Section VI and conclude the paper in Section VII.

## II. System Model

In this paper, we consider a single hop data gathering cluster consists of a set of sensor nodes and a sink node. The sensor nodes need to transmit data periodically to the sink node. When data transmission is completed in a collecting period, the sensor nodes all switch to sleep mode to save energy. Specifically as shown in Fig. 1, $N$ sensor nodes and a sink node form a Gaussian multi-access channel. Assume the collecting period is $T$. In a collecting period, sensor node $i$ holds a queue of $b_i$ packets and transmits its messages with signal $X_i$. The transmission rate is $R_i$ and the transmit power is $P_i$. Here, a packet is a collection of user data with fixed length of $B$ bits. Further, a packet can be divided into infinitesimal parts and encoded with different rates. The multi-access transmission is corrupted by a Additive White Gaussian Noise (AWGN) $Z$ distributed as Gaussian distribution $\mathcal{N}(0, \sigma^2)$. Without loss of generality, we assume that the channel gain from sensor node $i$ to the sink node is $f_i = 1$ for all $i \in I = \{1, 2, \ldots, N\}$. The received signal $Y$ at the sink node can be expressed as

$$Y = \sum_{i=1}^{N} X_i + Z. \tag{1}$$

Assume that all sensor nodes are with the same initial energy $E$, and a sensor network dies when one of the sensor nodes fails to complete its scheduled data transmission with the remaining energy. We define the lifetime $L$ of a sensor network as the number of collecting periods that the sensor network can accomplish. Merely decreasing the communication energy is inefficient to maximize the lifetime of a sensor network. Fair distribution of the energy consumption among sensor nodes should also be considered. In this paper, we investigate the min-max fair power scheduling in multi-access transmission with the objective to prolong the lifetime of the wireless sensor network.

### III. The Polymatroid Structure of the Multi-access Capacity Region and Power Region

In this section, we review the polymatroid structure of the Gaussian multi-access channel.

#### A. Polymatroid and Contra-polymatroid

For any $\mathbf{X} = (X_1, \ldots, X_N) \in \mathbb{R}_+^N$, let

$$\mathbf{X}(A) \triangleq \sum_{i \in A} X_i, \forall A \subseteq I \tag{2}$$

*Definition 1* Let set function $\rho : 2^I \to \mathbb{R}_+$ satisfy

1) $\rho(\emptyset) = 0$,
2) $\rho(A) \leq \rho(B), A \subseteq B \subseteq I$,
3) $\rho(A) + \rho(B) \geq \rho(A \cup B) + \rho(A \cap B)$.

In other words, $\rho$ is a monotone nondecreasing submodular function with $\rho(\emptyset) = 0$. Then, the polyhedron

$$(I, \rho) = \left\{ \mathbf{X} \in \mathbb{R}_+^N | \mathbf{X}(A) \leq \rho(A), \forall A \subseteq I \right\} \tag{3}$$

is called a polymatroid, where $I$ is called the ground set, and $\rho$ is the rank function of the polymatroid.

Correspondingly, we have the definition of the contra-polymatroid.

*Definition 2* Let set function $\rho : 2^I \to \mathbb{R}_+$ satisfy

1) $\rho(\emptyset) = 0$,
2) $\rho(A) \le \rho(B), A \subseteq B \subseteq I$,
3) $\rho(A) + \rho(B) \le \rho(A \cup B) + \rho(A \cap B)$.

Then, the polyhedron

$$(I, \rho) = \left\{ \mathbf{X} \in \mathbb{R}_+^N | \mathbf{X}(A) \ge \rho(A), \forall A \subseteq I \right\} \quad (4)$$

is called a contra-polymatroid. However, $\rho$ now is a supermodular function.

A vector $\mathbf{X} \in \mathbb{R}_+^N$ is called an independent vector of the contra-polymatroid $(I, \rho)$ if it is contained in the polyhedron defined by Eq. (4). For any $\mathbf{X}$ and $\mathbf{Y}$ in $\mathbb{R}_+^N$, let a partial order relation $\preceq$ be defined by

$$X \preceq Y \Leftrightarrow X_i \le Y_i, \forall i \in I. \quad (5)$$

A base of the contra-polymatroid $(I, \rho)$ is an independent vector which is minimal with respect to the partial order relation $\preceq$. In addition, the set containing all bases is called the dominant face of the contra-polymatroid $(I, \rho)$. The dominant face is a convex hull of extreme points of the contra-polymatroid, and these extreme points are called the vertices. The reader can refer to [4], [17] for further reference of polymatroids and submodular functions.

*B. The Polymatroid Structure of the Multi-access Capacity and Power Region*

The capacity region of a Gaussian multi-access channel possesses a polymatroid structure. Given a power vector $\mathbf{P} = (P_1, ..., P_N)$, the capacity polymatroid is

$$C_{MAC}(\mathbf{P}, \sigma^2) = \left\{ \mathbf{R} \in \mathbb{R}_+^N \middle| \mathbf{R}(A) \le \frac{1}{2} \log\left(1 + \frac{\mathbf{P}(A)}{\sigma^2}\right), A \subset I \right\}, \quad (6)$$

where $\sigma^2$ is the Gaussian noise power. Essentially, the capacity polymatroid imposes $2^N$ constraints on achievable rate vectors.

Given a rate vector $\mathbf{R} = (R_1, ..., R_N)$, the power vector $\mathbf{P}$ feasible to realize $\mathbf{R}$ in a Gaussian multi-access transmission forms a contra-polymatroid.

$$C_{MAC}(\mathbf{R}, \sigma^2) = \left\{ \mathbf{P} \in \mathbb{R}_+^N \middle| \mathbf{P}(A) \ge \sigma^2 \left(2^{2\mathbf{R}(A)} - 1\right), A \subset I \right\}. \quad (7)$$

Essentially, the power contra-polymatroid imposes $2^N$ constraints on the feasible transmit power vectors.

Due to the properties of the polymatroid and the contra-polymatroid, the rate scheduling and the power scheduling in a multi-access transmission are coupled and should be optimized cooperatively.

Denote the dominant face of $C_{MAC}(\mathbf{R}, \sigma^2)$ as $\mathcal{K}_{MAC}(\mathbf{R}, \sigma^2)$, the hyperplane that contains $\mathcal{K}_{MAC}(\mathbf{R}, \sigma^2)$ as $\mathcal{S}$. $\mathcal{K}_{MAC}(\mathbf{R}, \sigma^2)$ is the convex hull of vertices $\mathbf{V}_{\pi_1}, ..., \mathbf{V}_{\pi_{N!}}$. The power vector at a vertex $\mathbf{V}_{\pi_k}$ is determined by the successive decoding order $\pi_k$ which is a permutation on set $I$. There are $N!$ different permutations on set $I$ which correspond to $N!$ different successive decoding orders as well as $N!$ different vertices of the power contra-polymatroid. Adjusting the decoding order, we can achieve the power vector at any vertex. Denote the power vector at a vertex $\mathbf{V}_{\pi_k}$ as $(P_{\pi_k^{-1}(1)}, ..., P_{\pi_k^{-1}(N)})$, and its entries can be computed as

$$P_{\pi_k(i)} = \sigma^2 \left\{ 2^{2(\sum_{j=1}^{i} R_{\pi_k(j)})} - 1 \right\} - \sigma^2 \left\{ 2^{2(\sum_{j=1}^{i-1} R_{\pi_k(j)})} - 1 \right\}. \tag{8}$$

Correspondingly, the decoding order at the receiver should be the inverse order of $\pi_k$, i.e. $\pi_k(N) \to \pi_k(N-1) \to, ..., \to \pi_k(1)$. When $\mathbf{P} \in \mathcal{K}_{MAC}(\mathbf{R}, \sigma^2)$,

$$P_{sum} = \sum_{i=1}^{N} P_i = \sigma^2 \left\{ 2^{2(\sum_{i=1}^{N} R_i)} - 1 \right\}. \tag{9}$$

Namely, the sum power of any base in the dominant face $\mathcal{K}_{MAC}(\mathbf{R}, \sigma^2)$ is equal to $P_{sum}$. Therefore, the hyperplane $\mathcal{S}$ can be denoted as $\mathcal{S} = \left\{ \mathbf{P} \middle| P_{sum} = \sum_{i=1}^{N} P_i \right\}$.

Parallel to that in Section III-A, any power vector in the power contra-polymatroid is dominated (with respect to the partial order relation) by a base in the dominant face. Hence, it suffices to restrict our attention to the dominant face $\mathcal{K}_{MAC}(\mathbf{R}, \sigma^2)$.

*1) Realization of an Arbitrary Base in the Dominant Face:* Since the dominant face is a convex hull of the vertices, time sharing transmission among the vertices realizes any base in the dominant face $\mathcal{K}_{MAC}(\mathbf{R}, \sigma^2)$. Given the vertices of the contra-polymatroid, $\mathbf{V}_{\pi_1}, ..., \mathbf{V}_{\pi_{N!}}$, any power vector in the dominant face can be denoted as

$$\mathbf{P} = \mathbf{V}\beta, \tag{10}$$

where $\mathbf{V} = (\mathbf{V}_{\pi_1}, ..., \mathbf{V}_{\pi_{N!}})$ is the matrix formed by combining the vertices and $\beta = (\beta_1, ..., \beta_{N!})$ is the time sharing coefficient vector, where $\beta_k \geq 0, k = 1, ..., N!$ and $\sum_{k=1}^{N!} \beta_k = 1$. Specifically, to realize a base $\mathbf{P}$ in the dominant face, we first compute the corresponding time sharing coefficients vector $\beta$. Then, we divide the transmission duration $T$ into a serial of epochs $\xi_k, k \in \{1, 2, ..., K\}$. As there are $N!$ vertices, the number of the epochs is at most $K = N!$. During epoch $\xi_k$, $\xi_k = T\beta_k$, the sensor nodes transmit with the power vector at the vertex $\mathbf{V}_{\pi_k}$. At the sink node, the successive decoding order is the inverse order of $\pi_k$, i.e $\pi_k(N) \to \pi_k(N-1) \to ..., \to \pi_k(1)$. At the end of the time sharing transmission, a power scheduling strategy $\mathbf{P} = \sum_{k=1}^{N!} \beta_k \mathbf{V}_{\pi_k}$ is realized.

*2) Optimization in the Multi-access Capacity Polymatroid and Power Contra-polymatroid:*
Given $\lambda = (\lambda_1, ..., \lambda_N)$ and $\theta = (\theta_1, ..., \theta_N)$, the optimization problems

$$\max \lambda \cdot \mathbf{R}$$
$$s.t. \mathbf{R} \in C_{MAC}(\mathbf{P}, \sigma^2) \tag{11}$$

and

$$\min \theta \cdot \mathbf{P}$$
$$s.t. \mathbf{P} \in C_{MAC}(\mathbf{R}, \sigma^2) \tag{12}$$

are all proved to be optimally attained at some vertex of the corresponding polymatroid or contra-polymatroid [14]. Specifically, for the optimization problem in Eq. (12), the optimal solution is attained at a vertex $\mathbf{V}_{\pi_*}$ where $\pi_*$ is any permutation such that $\theta_{\pi_*(1)} \geq \theta_{\pi_*(2)} \geq , ..., \geq \theta_{\pi_*(N)}$.

Hereafter, we use an independent power vector and a point in the contra-polymatroid interchangeably and treat them as equivalent to a power allocation strategy.

## IV. An Optimal Scheduling to Minimize the Total Energy Cost

We refer the minicost transmission as an optimal scheduling that minimizes the total energy cost of the data gathering process. As shown in Fig. 2, we divide the minicost transmission into epochs $\xi_j$, $j \in J = \{1, ..., W\}$. In each epoch, the sensor nodes transmit with constant rates and powers. In other words, the transmit rate vector and the successive decoding order in each epoch are fixed. Denote the rate of sensor node $i$ in epoch $\xi_j$ as $r_{i,j}$, the successive decoding order in epoch $\xi_j$ as $\pi_j$, i.e. in epoch $\xi_j$ the decoding order is $\pi_j(N) \to \pi_j(N-1) \to ..., \to \pi_j(1)$.

Due to the convexity of Shannon channel capacity equation, the energy cost in p2p transmission can be reduced by lowering the transmission rate. In addition, it is proved in [13] that in a Gaussian multi-access transmission, if the rates of users lie in the dominant face, the users can be replaced by a super user transmitting at the sum rate with the sum power. Therefore, we argue that for the minicost transmission the sum rate of the sensor nodes must keep at the average rate $\frac{\sum_{i=1}^{N} b_i}{T}$ throughout the collecting period $T$.

*Theorem I:* It is sufficient that the sensor nodes transmit at a average rate throughout the collecting period to minimize the total energy cost in the data gathering. Specifically, for sensor node $i$, its rate keeps at $r_{i,j} = \frac{b_i}{T}$, $\forall j \in J$.

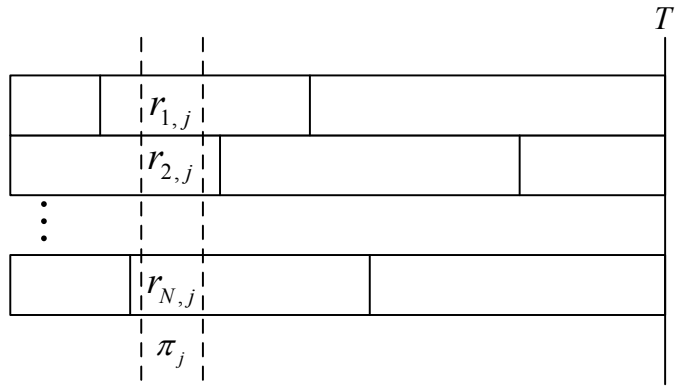

Fig. 2. The minicost transmission process is divided into a series of epochs.

*Proof:* It is sufficient to achieve the minicost transmission if, at any epoch $j$, $\sum_{i=1}^{N} r_{i,j} = \frac{\sum_{i=1}^{N} b_i}{T}$. When senor node $i$ transmits at a average rate $r_{i,j} = \frac{b_i}{T}$, $\forall j \in J$, the sum rate keeps at $\frac{\sum_{i=1}^{N} b_i}{T}$ throughout the transmission. Therefore, it is sufficient that the minicost transmission is achieved. □

*Theorem II:* Given the transmission rate vector, the optimal power scheduling, which achieves the minicost transmission, corresponds to successive decoding by a fixed order $\pi_j = \pi$, $\forall j \in J$, throughout the collecting period.

*Proof:* Given the transmission rate vector $\mathbf{r}_j = (r_{1,j}, ..., r_{N,j})$ in epoch $\xi_j$, the optimal power control which minimizes the sum power cost corresponds to successive decoding by a decreasing order of the channel gains [14]. In a symmetric multi-access channel, the channel gains are same for the sensor nodes. Therefore, any fixed successive decoding order $\pi$ throughout the collecting period is sufficient to achieve the minicost transmission. □

Given the average rate vector $\mathbf{R} = (\frac{b_1}{T}, ..., \frac{b_N}{T})$, successive decoding according to any fixed order $\pi$ is enough to realize the minicost transmission. In other words, the sensor nodes may transmit with fixed power vector for all epochs. In the following analysis for the min-max fair power scheduling, we also take the average rate vector $\mathbf{R} = (\frac{b_1}{T}, ..., \frac{b_N}{T})$ as the transmission rate vector for the sensor nodes. However, as we will show, the min-max fair power scheduling demands a time sharing transmission among different vertex power vectors. Hence, different successive decoding orders are required in different epochs.

## V. A Min-max Fair Scheduling in the Multi-access Power Contra-polymatroid

To balance the energy consumption among sensor nodes, we study the fair power scheduling in the multi-access transmission. Specifically, we investigate the min-max power allocation in a multi-access power contra-polymatroid. The min-max fairness is defined as below.

*Definition 3 (Min-max fairness).* A feasible power allocation vector **P** is min-max optimal if and only if a decrease of any power within the domain of feasible power allocation region must be at the cost of an increase of some already higher or equal power. Formally, for any other feasible power allocation vector **P**′, if $P'_i < P_i$, then there must exist some $j$ such that $P_j \geq P_i$ and $P'_j \geq P_j$.

With the average rate vector adopted as the transmission rate vector, the optimization model of the min-max scheduling is denoted as

$$\text{min-max } \mathbf{P}$$
$$s.t. \mathbf{P} \in \mathcal{K}_{MAC}(\mathbf{R}, \sigma^2) \tag{13}$$

where, $\mathbf{R} = (\frac{b_1}{T}, ..., \frac{b_N}{T})$ is the average rate vector, and $\mathbf{P} \in \mathcal{K}_{MAC}(\mathbf{R}, \sigma^2)$ means that the power vector must conform to the constraints of the multi-access power contra-polymatroid determined by **R** and the Gaussian noise power $\sigma^2$.

## A. Min-max Fair and Lexicographically Optimal

In this part, we briefly review the properties of the min-max fairness (dual to the max-min fairness) discussed in [3], [8], [10].

Assume an order mapping $\mathcal{T} : \mathbb{R}^N_+ \to \mathbb{R}^N_+$ that sorts the elements of a vector **X** in a decreasing order.

$$\mathcal{T}(X_1, ..., X_N) = (X_{(1)}, ..., X_{(N)})$$
$$X_{(1)} \geq ... \geq X_{(N)} \tag{14}$$

where, $X_{(i)}$, $i = 1, ..., N$, is an element of **X**.

*Definition 5( [10]).* For vector **X** and **Y** in the feasible region $\mathcal{X}$, **X** is lexicographically smaller than or equal to **Y**, denoted as $\mathbf{X} \leq^{lex} \mathbf{Y}$, if for some $i \in I$,

$$X_j = Y_j, j < i$$
$$X_i < Y_i$$

or

$$X_i = Y_i, \forall i \in I$$

where, $X_i$ and $Y_i$ denote the $i$th element of vector **X** and **Y**, respectively.

*Definition 6( [3]).* If **X** is lexicographically optimal in the feasible region $\mathcal{X}$, then

$$\mathcal{T}(\mathbf{X}) \leq^{lex} \mathcal{T}(\mathbf{Y}), \forall \mathbf{Y} \in \mathcal{X} \tag{15}$$

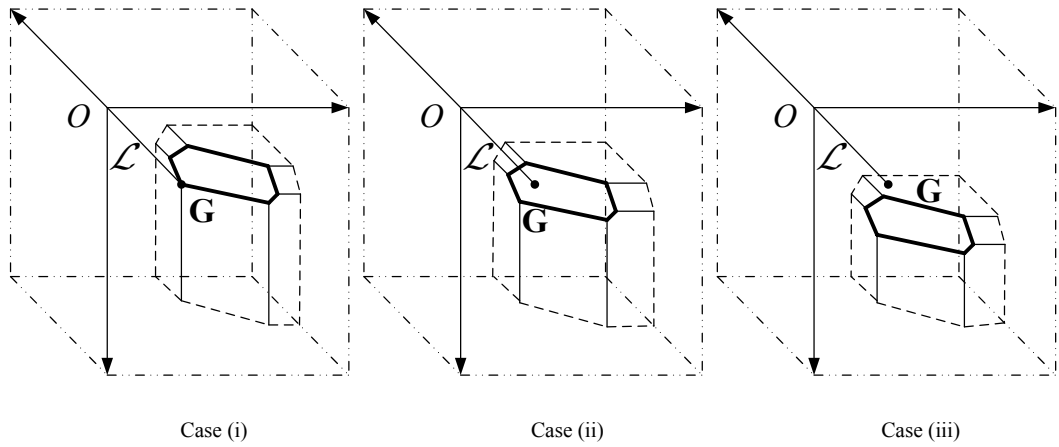

Fig. 3. Three cases of the min-max scheduling in a $N = 3$ sensor network. The hexagon circled by thick lines denotes the dominant face $\mathcal{K}_{MAC}(\mathbf{R}, \sigma^2)$. Note that the power contra-polymatroid is not blocked by the dashed lines. It actually extends to infinity along its six boundary facets.

It was proved that if a min-max optimal vector exists on set $\mathcal{X}$, then it is unique. Besides, the min-max optimal vector is the unique lexicographically optimal vector. If the feasible set $\mathcal{X}$ is convex and compact, then the unique lexicographically optimal vector is the min-max optimal vector. Therefore, in a convex compact set, the problem searching for the min-max optimal vector is equivalent to the problem searching for the lexicographically optimal vector. Because the dominant face of a multi-access power contra-polymatroid is a convex and compact set, the lexicographically optimal vector can be used instead to achieve the min-max fairness. In the following part, we will prove that the lexicographically optimal vector in the dominant face is the vector with minimal distance from the equal allocation vector.

## B. Geometry Analysis

We first relax the constraint in Eq.(13) and search for the min-max optimal power vector on the hyperplane $\mathcal{S}$. The optimization problem then can be expressed as

$$\text{min-max } \mathbf{P}$$
$$s.t. P_{sum} = \sum_{i=1}^{N} P_i \tag{16}$$

The solution to Eq.(16) is the solution to the following equations,

$$\begin{cases} P_1 = P_2 =, ..., = P_N \text{ (Line } \mathcal{L}) \\ P_{sum} = \sum_{i=1}^{N} P_i \text{ (Hyperplane } \mathcal{S}) \end{cases} \tag{17}$$

The solution is the point where line $\mathcal{L}$ intersects with hyperplane $\mathcal{S}$. We denote the intersection point as $\mathbf{G}$, then $\mathbf{G} = \left(\frac{P_{sum}}{N}, ..., \frac{P_{sum}}{N}\right)$.

Let us return to the optimization problem in Eq.(13), and constrain the feasible region to the dominant face $\mathcal{K}_{MAC}(\mathbf{R}, \sigma^2)$. According to different geometric relationship between the equal power point $\mathbf{G}$ and $\mathcal{K}_{MAC}(\mathbf{R}, \sigma^2)$, the optimization problem in Eq.(13) consists of three cases.

(i) $\exists k, k \in \{1, ..., N!\}, \mathbf{G} = \mathbf{V}_{\pi_k}$. The equal power point $\mathbf{G}$ coincides with one of the vertices. In this case, point $\mathbf{G}$ is the min-max optimal point that we are searching for. To realize the min-max power allocation, the sensor nodes should transmit with the equal power vector. At the sink node, successive decoding should be performed according to the inverse order of $\pi_k$.

(ii) $\mathbf{G} \in \mathcal{K}_{MAC}(\mathbf{R}, \sigma^2)$ and $\forall k \in \{1, ..., N!\}, \mathbf{G} \neq \mathbf{V}_{\pi_k}$. Under this condition, the min-max power point is still the equal power point $\mathbf{G}$. However, as $\mathbf{G}$ is not one of the vertices, time sharing among the vertices is indispensable to realize the equal power vector. In other words, the sensor nodes have to transmit according to different vertices in different epochs.

(iii) $\mathbf{G} \in \mathcal{K}_{MAC}^c(\mathbf{R}, \sigma^2)$. The intersection point $\mathbf{G}$ is out of the dominant face $\mathcal{K}_{MAC}(\mathbf{R}, \sigma^2)$. In this situation, the equal power point $\mathbf{G}$ is no longer feasible.

The three cases of the min-max scheduling in an $N = 3$ sensor network are shown in Fig. 3.

Denote the distance $d$ between points $\mathbf{X}$ and $\mathbf{Y}$ in the $N$-dimensional space as

$$d(\mathbf{X}, \mathbf{Y}) = \|\mathbf{X}, \mathbf{Y}\|_2^2 = \sum_{i=1}^{N}(X_i - Y_i)^2. \tag{18}$$

We will prove that in Case (iii) the min-max optimal base in the dominant face $\mathcal{K}_{MAC}(\mathbf{R}, \sigma^2)$ is the base with the minimal distance to the equal power point $\mathbf{G}$.

First, we review a few definitions and results in [3] and we paraphrase them as below.

*Definition 6* $\forall \mathbf{X} \in (I, \rho)$, the saturated set of $\mathbf{X}$, denoted as $sat(\mathbf{X})$, is the set of all elements $i \in I$ such that for any $e > 0$ the vector $\mathbf{Y} \in \mathbb{R}_+^N$ defined by

$$\mathbf{Y} = \mathbf{X} - e\mathbf{U}_i, \mathbf{U}_i = (0, ..., 1_{ith}, ..., 0) \tag{19}$$

is out of the contra-polymatroid $(I, \rho)$. Here, $\mathbf{U}_i$ denotes the unit vector with the element in the $i$th dimension $U_{i,i} = 1$. If $\mathbf{X}$ is on the boundary of the contra-polymatroid, $sat(\mathbf{X}) \neq \emptyset$. If $\mathbf{X}$ is in the dominant face, $sat(\mathbf{X}) = I$.

*Lemma 1* $\forall \mathbf{X} \in (I, \rho)$, the set $A = sat(\mathbf{X})$ satisfies

$$\mathbf{X}(A) = \rho(A). \tag{20}$$

Moreover, let $\mathfrak{a}_1$ be defined by

$$\mathfrak{a}_1 = \{A | A \subseteq I, \mathbf{X}(A) = \rho(A)\}. \tag{21}$$

Then $\mathfrak{a}_1$ is a distributive lattice with partial order relation defined by set inclusion and $sat(\mathbf{X})$ is the maximum element of $\mathfrak{a}_1$.

*Definition 7* $\forall \mathbf{X} \in (I, \rho)$, if $i \in sat(\mathbf{X})$, the dependent set of $\mathbf{X}$ with respect to $i$, denoted as $dep(\mathbf{X}, i)$, is the set of all elements $j \in I$ such that for some $e > 0$ the vector $\mathbf{Y} \in \mathbb{R}_+^N$ defined by

$$\mathbf{Y} = \mathbf{X} - e\mathbf{U}_i + e\mathbf{U}_j, \mathbf{U}_i = (0, ..., 1_{ith}, ..., 0), \mathbf{U}_j = (0, ..., 1_{jth}, ..., 0) \tag{22}$$

is in the contra-polymatroid $(I, \rho)$. If $i$ is not in the set $sat(\mathbf{X})$, then define $dep(\mathbf{X}, i) = \emptyset$. $\forall i \in I$, if $dep(\mathbf{X}, i) \neq \emptyset$, then $i \in dep(\mathbf{X}, i)$.

*Lemma 2* $\forall \mathbf{X} \in (I, \rho)$, $sat(\mathbf{X}) \neq \emptyset$, if for $i \in sat(\mathbf{X})$, $dep(\mathbf{X}, i) = B$, then

$$\mathbf{X}(B) = \rho(B). \tag{23}$$

Moreover, let $\mathfrak{a}_2$ be defined by

$$\mathfrak{a}_2 = \{B | B \subseteq I, i \in B, \mathbf{X}(B) = \rho(B)\}. \tag{24}$$

Then $\mathfrak{a}_2$ is a distributive lattice with partial order relation defined by set inclusion and $dep(\mathbf{X}, i)$ is the minimum element of $\mathfrak{a}_2$.

Now, we start to prove that the min-max optimal vector is the base with minimal distance from the equal allocation point $\mathbf{G}$. We follow basically the same line with the process in [3]. However, the obtained corollaries reveal the geometry property of the min-max fairness in a power contra-polymatroid. In addition, they suggest a convenient method to compute the min-max optimal vector and the associated time sharing coefficients.

*Lemma 3* Let $\mathbf{X}$ be a base of the contra-polymatroid $(I, \rho)$ and the distinct numbers of $\mathbf{X}$ be given by $C_1 > C_2 > ... > C_p$. Furthermore, define $S_j \subseteq I$, $j = 1, 2, ..., p$ by

$$S_j = \left\{l | l \in I, X_l \geq C_j\right\}, j = 1, 2, ..., p. \tag{25}$$

The following two conditions are equivalent:

1) $\mathbf{X}$ is the lexicographically optimal base of $(I, \rho)$.
2) $\emptyset \neq dep(\mathbf{X}, i) \subseteq S_j$, $i \in S_j$, $j = 1, 2, ..., p$.

*Proof.* If 1) is true, 2) follows from the definition of lexicographically optimal. Because the dominant face is a compact convex set, the lexicographically optimal base is the min-max optimal point in the dominant face. If 2) is true, 1) follows from the definition of min-max fairness. □

*Theorem III* Let $\mathbf{X}^*$ be the lexicographically optimal base of $(I, \rho)$ and $\mathbf{X}'$ be the unique optimal solution to the following optimization problem,

$$\min f(\mathbf{X}) = \tfrac{1}{2} \sum_{i=1}^{N} X_i^2 \qquad (26)$$

$$s.t. \mathbf{X} \text{ is a base of the contra-polymatroid } (I, \rho)$$

then, we have $\mathbf{X}^* = \mathbf{X}'$.

*Proof.* Let $\mathbf{X}'$ be the unique optimal base to the optimization problem and the distinct numbers of $\mathbf{X}'$ be given by $C_1 > C_2 > ... > C_p$. Then, we have

$$\frac{\partial}{\partial X_i} f(\mathbf{X})|_{\mathbf{X}=\mathbf{X}'} = X_i'. \qquad (27)$$

Therefore, $\forall i, j \in I$, if $X_i' > X_j'$, we have

$$j \notin dep(\mathbf{X}', i), \qquad (28)$$

since otherwise, there would exist a base which yields a smaller value of $f$ than $\mathbf{X}'$. Consequently, for $S_j \subset I$, $j = 1, 2, ..., p$,

$$\emptyset \neq dep(\mathbf{X}', i) \subseteq S_j = \{l | l \in I, X_l' \geq C_j\}, \forall i \in S_j. \qquad (29)$$

It follows from Eq. (29) and lemma 3 that $\mathbf{X}'$ coincides with the unique lexicographically optimal base $\mathbf{X}^*$ of $(I, \rho)$. □

*Corollary 1* $\forall \mathbf{X}, \mathbf{Y} \in \mathbb{R}_+^N$, define the distance between $\mathbf{X}$ and $\mathbf{Y}$ as Eq. (18). $\forall a \in \mathbb{R}_+$, define the equal allocation vector as $\mathbf{V}_a = (a, ..., a)$. Let $\mathbf{X}^*$ be the lexicographically optimal base of $(I, \rho)$, then we have

$$d(\mathbf{X}^*, \mathbf{V}_a) \leq d(\mathbf{X}, \mathbf{V}_a), \qquad (30)$$

where, $\mathbf{X}$ is an arbitrary base of the contra-polymatroid $(I, \rho)$.

*Proof.* For any base $\mathbf{X}$ of the contra-polymatroid $(I, \rho)$,

$$\begin{aligned} &d(\mathbf{X}, \mathbf{V}_a) \\ &= \sum_{i=1}^{N}(X_i - V_{a,i})^2 \\ &= \sum_{i=1}^{N} X_i^2 + V_{a,i}^2 - 2X_i V_{a,i} \\ &= Na^2 + \sum_{i=1}^{N} X_i^2 - 2a \sum_{i=1}^{N} X_i. \end{aligned} \qquad (31)$$

It follows from that $\sum_{i=1}^{N} X_i$ is same for all bases and Theorem III that the lexicographically optimal base $\mathbf{X}^*$ is the base with the minimal distance to the equal allocation vector $\mathbf{V}_a$. □

*Corollary 2* Let $\mathbf{X}^*$ be the lexicographically optimal base of $(I, \rho)$, then it is also the lexicographically optimal independent vector of the contra-polymatroid $(I, \rho)$.

*Proof.* Because a base is an independent vector minimal with respect to the partial order relation $\preceq$, the lexicographically optimal base is also the lexicographically optimal independent vector of the contra-polymatroid $(I, \rho)$. □

Because the dominant face of the power contra-polymatroid is a compact convex set, the unique lexicographically optimal base is the min-max optimal power vector for which we are searching. Since in Case (iii) the min-max power point locates itself at the point with the minimal distance to the equal power point $\mathbf{G}$ and in Case (i) and Case (ii) the min-max optimal point is $\mathbf{G}$ itself, we can conclude that the min-max optimal point in the power contra-polymatroid $C_{MAC}(\mathbf{R}, \sigma^2)$ is the base with the minimal distance to the equal power point $\mathbf{G}$.

*C. Computing the Time Sharing Coefficients to Achieve the Min-max Optimal Power Scheduling*

Following Corollary 1, the min-max optimal vector $\mathbf{P}^*$ can be denoted as

$$\mathbf{P}^* = \arg \min_{\mathbf{P} \in \mathcal{K}(\mathbf{R}, \sigma^2)} d(\mathbf{P}, \mathbf{G}) \tag{32}$$

As discussed in Section III, any power vector $\mathbf{P}$ in the dominant face can be denoted as $\mathbf{P} = \mathbf{V}\beta$. Therefore, the problem in Eq.(32) can be converted to an optimization problem as below.

$$\min \|\mathbf{V}\beta, \mathbf{G}\|_2^2$$
$$s.t. \sum_{k=1}^{N!} \beta_k = 1 \tag{33}$$
$$\beta_k \geq 0, k \in \{1, ..., N!\}$$

The optimization function $\|\mathbf{V}\beta, \mathbf{G}\|_2^2 = \beta^T \mathbf{V}^T \mathbf{V} \beta - 2\mathbf{G}^T \mathbf{V} \beta + \mathbf{G}^T \mathbf{G}$ is a convex quadratic function and the corresponding feasible region is a convex set. There is a number of existing algorithms to solve this convex quadratic optimization problem and compute the time sharing coefficients vector $\beta$. Note that multiple optimal solutions may exist for the optimization problem in Eq. (33).

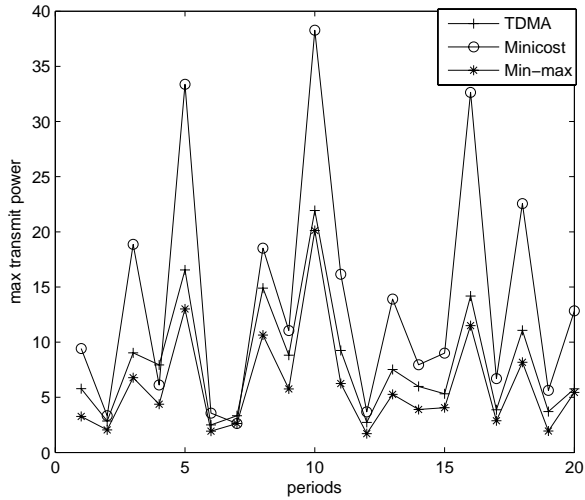

Fig. 4. The comparison of maximum transmit power among sensor nodes.

## VI. SIMULATION RESULTS

To validate the proposed method for min-max fair scheduling, we apply it to a multi-access sensor network in the simulations. Specifically, we consider a WSN with 4 sensor nodes and a sink node. The sensor nodes are with the same initial energy $E = 2$J. We compare the min-max fair scheduling with the minicost as well as the optimal TDMA scheduling. In the optimal TDMA scheduling, the transmission time is allocated optimally according to the backlogs at the sensor nodes. In the comparison, parameter $\lambda$ denotes the highest packet arriving rate at sensor nodes. Given $\lambda$, the backlog lengthes at the sensor nodes are distributed uniformly between $(0, \lambda]$ in a collecting period. We assume a packet length of $B = 30$bit and a collecting period of $T = 30$S, with the Gaussian noise power $\sigma^2 = -30$dB.

In Fig. 4, we compare the maximum transmit power among sensor nodes for different power scheduling strategies. For the TDMA transmission, the transmit powers of the sensor nodes are normalized to the entire collecting period $T$. With $\lambda = 1$, Fig. 4 shows that the min-max optimal scheduling minimizes the maximum transmit power. For a fixed collecting period $T$, the min-max optimal scheduling also consumes the least maximum energy among the sensor nodes. Compared to TDMA or minicost transmission, it distributes the total energy cost more fairly among sensor nodes.

With $\lambda = 1$, Fig. 5 shows the comparison of the sum energy cost for different power scheduling strategies. Since we assume a symmetric channel model, the sum power cost are same for TDMA, minicost and the min-max optimal scheduling. For a fixed collecting period $T$, the sum energy cost are also same for these three scheduling strategies. Combining this result with that obtained from Fig. 4, we conclude that in a symmetric multi-access transmission, the min-max fair scheduling effectively balances the energy consumption among the sensor nodes without increasing the sum energy cost.

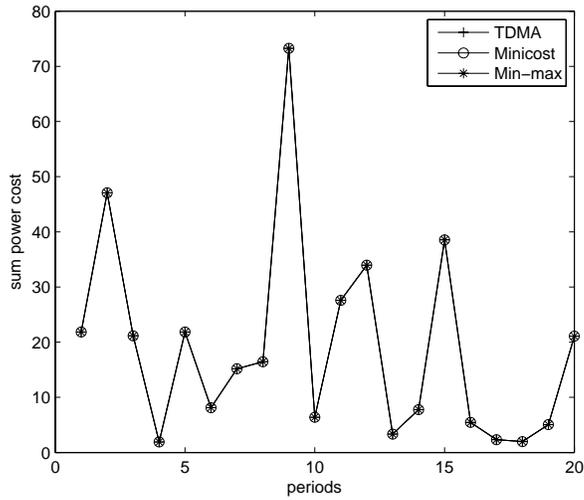

Fig. 5. The comparison of sum energy consumption in a collecting period.

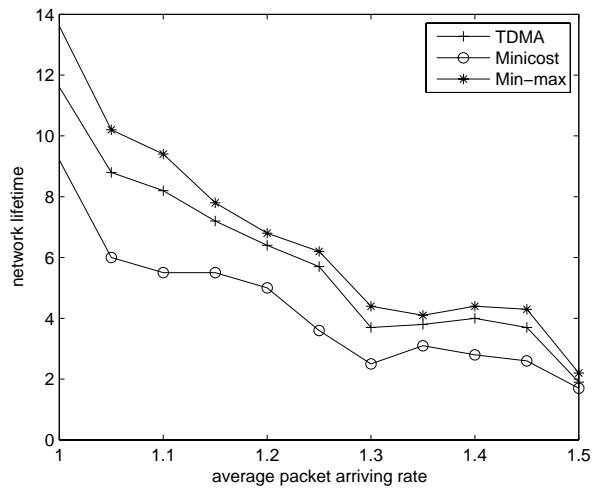

Fig. 6. The comparison of network lifetime under different power scheduling strategies.

With the highest packet arriving rate $\lambda$ increasing, the network lifetimes for all the three scheduling strategies decrease. However, at any fixed $\lambda$, the min-max optimal scheduling 17outperforms the TDMA as well as the minicost scheduling on prolonging the network lifetime. As it is shown in Fig. 6, when $\lambda = 1.0$, the network lifetime under the minicost scheduling is 9.2 periods. While, the network lifetime under the min-max optimal scheduling is 13.6 periods, which exceeds that of the minicost transmission about 50%.

## VII. Concluding Remarks

In this paper, we prove that the min-max optimal vector in a contra-polymatroid is the base with the minimal distance to the equal allocation vector. Since time sharing among the vertices can realize any base in the contra-polymatroid, the min-max scheduling problem evolves as a convex optimization problem computing the time sharing coefficients. To compute the min-max scheduling strategy, the proposed method is superior to the traditional algorithms in computer networking, as it interprets the geometry property of the min-max fairness in a contra-polymatroid and can solve the min-max scheduling problem with any existing convex optimization algorithms. Moreover, it can be extended in the following two aspects.

Although we consider a symmetric multi-access channel in this paper, it is easily to extend the obtained results to an asymmetric channel model. Redefine the distance $d$ between points $\mathbf{X}$ and $\mathbf{Y}$ in the $N$-dimensional space as

$$d(\mathbf{X}, \mathbf{Y}) = \|\mathbf{X}, \mathbf{Y}\|_2^2 = \sum_{i=1}^{N} f_i (X_i - Y_i)^2 \tag{34}$$

where, $f_i$ is the channel gain of sensor node $i$. Following the same line as in Section V, we can prove that the min-max optimal point in the power contra-polymatroid $C_{MAC}(\mathbf{R}, \sigma^2)$,

$$C_{MAC}(\mathbf{R}, \sigma^2) = \left\{ \mathbf{P} \in \mathbb{R}_+^N \,\middle|\, \mathbf{Q}(A) \geq \sigma^2 \left( 2^{2\mathbf{R}(A)} - 1 \right), A \subset I \right\}, \tag{35}$$

is the base with the minimal distance to the equal power point $\mathbf{G}$. Here, $\mathbf{Q} = (f_1 P_1, ..., f_N P_N)$ is the received power vector. Essentially, the channel gain vector $\mathbf{f} = (f_1, ..., f_N)$ can be considered as a weight vector. When the sensor nodes are with different initial energy or different transmission workload, we apply a parallel weight vector to achieve a weighted min-max fairness among the sensor nodes. By such weighted min-max fair scheduling, we can prolong the network lifetime of a heterogenous sensor network.

Dual to the min-max power scheduling considered in this paper, the max-min rate allocation problem in a capacity polymatroid can be solved in the same line. Analogically, we can prove that the max-min optimal rate vector in a multi-access capacity polymatroid is the base with the minimal distance to the equal allocation vector and obtain its associated time sharing

January 25, 2010      DRAFT

coefficients by convex optimization algorithms. By time sharing transmission according to the obtained coefficients, the max-min rate scheduling can be realized.

**Xiaomao MAO** received the B.Sc. degree in electronics engineering from Dalian University of Technology, China, in 2005. She is currently working toward the Ph.D. degree in communication and information systems from Department of Information Science and Electronic Engineering, Zhejiang University, China. Her research interests are in information theory and wireless communication, with a current focus on energy efficient transmission scheduling and sensor network design.

**Huifang CHEN** received the B.S. degree in electronic engineering, M.S, and Ph. D. degrees in communication and information systems from Zhejiang University, Hangzhou, China, in 1994, 1997, and 2000, respectively. Since 2000, she has been with Zhejiang University where she is now an associate professor in the department of Information Science and Electronic Engineering. From Oct. 2005 to Sept. 2007, she worked as a post-doctoral researcher supported by the Japanese Government Scholarship in the department of computer science, Shizuoka University, Japan. Her current research interests include information theory and coding, wireless networks and security issue in wireless networks. She is a member of the IEEE.

**Peiliang QIU** received his B.Sc degree in electronic engineering from Harbin Engineering University, China, in 1967 and M.S. degree in Information Science from the Chinese Academy of Science. He joined as a faculty in Zhejiang University in 1981, where he is currently a professor of Electronic Engineering. His research interests include information and coding theory, wireless communication, cooperative relay transmission and sensor network design. He has authored or co-authored over 200 academic papers and published 3 books. He serves as an editor for Journal of Communication, Journal of Zhejiang University. He is the vice president of Chinese Institute of Electronic, Information Theory Association.